\documentstyle[12pt,qqaajart]{article}

\author{Vadim L. Vereschagin\thanks{Talk on Conference of Young Scientists
of Moscow State University, February 1987}\\
Irkutsk Computing Center, P.O.Box 1233, Irkutsk 664033, RUSSIA}
\title{ON CLASSIFICATION OF DEFORMATIONS OF THE GARDNER BRACKET OVER LOCAL
INFINITESIMAL TRANSFORMATIONS
}
\input tcilatex

\QQQ{Language}{
American English
}

\begin{document}

\maketitle
Let $\Gamma $ be a Lie algebra of local functionals of scalar one-argument
function with commutator determined by the Gardner-Zakharov-Faddeev (GZF)
bracket \cite{teorsol} :

$$
\left\{ P(\Psi \},Q(\Psi )\right\} =\int \frac{\delta P}{\delta \Psi (x)}%
\partial _x\frac{\delta Q}{\delta \Psi (x)}dx
$$

Investigate the following problem: classify infinitesimal deformations of
the GZF bracket like

$$
\left\{ \Psi (x),\Psi (y)\right\} _\epsilon =\delta ^{\prime }(x-y)+\epsilon
\varphi _s\left( \Psi (x),...,\Psi ^{(N_s)}(x)\right) \delta ^{(s)}(x-y)
$$

\noindent (summation in repeating indices).

Consider a deformation trivial if it can be induced by an infinitesimal
Lie-B\"acklund transformation:

$$
\Psi (x)\rightarrow \Psi (x)+\epsilon F\left( \Psi (x),...,\Psi
^{(M)}(x)\right) \equiv \bar \Psi (x,\epsilon )
$$

$$
\left\{ \Psi (x),\Psi (y)\right\} _\epsilon =\left\{ \bar \Psi (x,\epsilon
),\bar \Psi (y,\epsilon )\right\} (mod\epsilon ^2),\quad where
$$

\begin{equation}
\label{111}\{,\}\ is\ the\ GZF\ bracket\
\end{equation}

This problem is important for investigation of perturbations of conservative
systems like the Korteweg-de Vries equation and the theory of averaging.

Define a complex of cochains on algebra $\Gamma $ in the following way:
polylinear skew-symmetric $n$-form $C^n$ on point functionals takes the
meaning

$$
C^n\left( \Psi (x_1),...,\Psi (x_n)\right) =a_{i_1,...,i_{n-1}}\left( \Psi
(x_1),...,\Psi ^{(P)}(x_1)\right) \delta ^{(i_1)}(x_1-x_2)...\delta
^{(i_{n-1})}(x_1-x_{n-1})
$$

\noindent where $a_{i_1,...,i_{n-1}}\left( \Psi (x_1),...,\Psi
^{(P)}(x_1)\right) $ is some function. The differential in the complex is
determined by the following formula:

$$
\begin{array}{c}
\delta C^n\left( \Psi (x_1),...,\Psi (x_{n+1})\right) =(-1)^s\delta
^{(i_1)}(x_1-x_2)...\delta ^{(j+1)}(x_1-x_s)...\delta
^{(i_{n-1})}(x_1-x_{n+1})\times \\
a_{i_1,...,i_{n-1}}\Psi ^{(j)}(x_1)+\partial _{x_1}^{j+1}\left[ \delta
(x_1-x_2)\delta ^{(i_1)}(x_1-x_3)\delta
^{(i_{n-1})}(x_1-x_{n+1})a_{i_1,...,i_{n-1}}\Psi ^{(j)}(x_1)\right]
\end{array}
$$

It is easy to verify that the operation $\delta $ is a restriction of
standard differential of the Lie algebras theory.

\underline{STATEMENT 1.} The second cohomology group of this complex yields
the needed classification of the bracket deformations.

Proof is based on direct verification of triviality of any 2-coborder in the
sense of (\ref{111}).

Investigation of 2-cocycles like $\varphi _s\delta ^{(s)}(x-y),\ s=0,1,...S$
(cocycle of order $S$) indicates that there are no nontrivial cocycles while
$S\leq 5$ (cf. \cite{asta} ).

\underline{HYPOTHESIS.} The second cohomology group of the complex specified
above is trivial (cf. with main hypothesis in \cite{astvin} ).

\underline{STATEMENT 2.} If for a cocycle of order $S$ one has $\varphi
_s=\varphi _s\left( \Psi ,...,\Psi ^{(N_s)}\right) ,\ s=0,1,2,...S,$ then $%
N_s=N(S)-s.$

The cocycle is obviously trivial under $N(S)<2S,$ and while $N(S)\geq 2S$
the dependence of $N(S)$ can be extracted from the equation

$$
\det \left( C_{N+1-j}^i+diag(-1)^{N+i+1}\right) =0;\quad i,j=0,1,...,S;\quad
S=1,3,5,...
$$

\underline{PROBLEM.} Roots of this equation are as follows: $%
2S,2S+1,...,3S-2,3S.$

\

\end{document}